# Cortical network reconfiguration aligns with shifts of basal ganglia and cerebellar influence

Kimberly Nestor (1-3)

1) Department of Psychology, Carnegie Mellon University, Pittsburgh PA, USA
2) Center for the Neural Basis of Cognition, Pittsburgh PA, USA
3) Carnegie Mellon Neuroscience Institute, Pittsburgh PA, USA

**Corresponding author:**
Kimberly Nestor
Email: kimberly.a.nestor@gmail.com

**ABSTRACT**

Mammalian functional architecture flexibly adapts, transitioning from integration where information is distributed across the cortex, to segregation where information is focal in densely connected communities of brain regions. This flexibility in cortical brain networks is hypothesized to be driven by control signals originating from subcortical pathways, with the basal ganglia shifting the cortex towards integrated processing states and the cerebellum towards segregated states. In a sample of healthy human participants (N=242), we used fMRI to measure temporal variation in global brain networks while participants performed two tasks with similar cognitive demands (Stroop and Multi-Source Inference Task (MSIT)). Using the modularity index, we determined cortical networks shifted from integration (low modularity) at rest to high modularity during easier i.e. congruent (segregation). Increased task difficulty (incongruent) resulted in lower modularity in comparison to the easier counterpart indicating more integration of the cortical network. Influence of basal ganglia and cerebellum was measured using eigenvector centrality. Results correlated with decreases and increases in cortical modularity respectively, with only the basal ganglia influence preceding cortical integration. Our results support the theory the basal ganglia shifts cortical networks to integrated states due to environmental demand. Cerebellar influence correlates with shifts to segregated cortical states, though may not play a causal role.

**INTRODUCTION**

The mammalian brain excels at rapidly adapting complex behavioral repertoires in response to, or anticipation of, environmental changes (Young et al., 2017; Wendelken et al., 2016b; Shine, Koyejo & Poldrack, 2016b). This behavioral versatility reflects a flexibility in the underlying functional architecture of the brain, particularly in the cerebral cortex (Cohen & D'Esposito, 2016; Braun et al., 2015; Allen et al., 2014; Betzel et al., 2022; Chen et al., 2016; Cole et al., 2014; Cole et al., 2013; Davison et al., 2015; Douw et al., 2016; Finc et al., 2017; Gallen et al., 2016; Geib et al., 2017; Gonzalez-Castillo et al., 2015; Guimerà & Nunes Amaral, 2005; Langer et al., 2012; Lynn et al., 2021; Mattar et al., 2015; Mišić et al., 2016; Patil et al., 2021; Schultz & Cole, 2016; Vatansever et al., 2017; Zalesky et al., 2014). Viewed from a network perspective (Sporns & Betzel, 2016; Sporns, 2013), where distinct brain regions form graph nodes and the mutual connectivity between them graph edges, the topology of macroscopic brain networks shows not only a high degree of individuality, but also flexibility in response to contextual changes (Gratton et al. 2018; Shine et al. 2016). Specifically, these networks rapidly shift between more integrated (i.e., greater communication between communities of nodes) and segregated (i.e., more modular organization of communities) states, depending on immediate cognitive demands (Sporns, & Betzel, 2016; Shine & Poldrack, 2018; Bassett et al., 2013; Bassett et al., 2015; Bertolero, Yeo & D'Esposito, 2015; Bertolero, Yeo, Bassett & D'Esposito, 2018; Betzel et al., 2016; Capouskova et al., 2023; Cohen & D'Esposito, 2016; Fair et al., 2007; Mohr et al., 2016; Shine, 2021; Shine et al., 2016; Tononi, Sporns & Edelman, 1994; Vatansever et al., 2015; Wang et al., 2021; Westphal, Wang & Rissman, 2017; Zippo et al., 2018).

More specifically, as immediate cognitive demand increases, macroscopic brain networks shift from more segregated states to more integrated states, reflecting the increased need for distributed computations and the need to integrate locally processed information. For example, in tasks requiring the identification of sequential versus randomly scrambled stories, functional macroscopic brain networks shift into more integrated states than during a more simplistic task of listening to randomly scrambled words (Owen et al., 2021; Bertolero, Yeo & D'Esposito, 2015; Cohen & D'Esposito, 2016). However,there is a trade off between integrative and segregative states in order to meet task demands (Schultz & Cole, 2016; Wang, et al., 2021) and minimize loss of metabolic resources (van den Heuvel, et al., 2012; Langer, et al., 2012). In this way, the global network architecture reflects the moment-by-moment information processing demands of the environment, while ensuring upon habituation to a task, the cortical network converges to a state that expends the least amount of energy for skill maintenance (Achard & Bullmore, 2007; Collin et al., 2014).

While the existence of these flexible network architectures in the cortex is fairly well established, precisely how they are implemented is unknown. Moreover,use of blood-oxygen level-dependent (BOLD) activity may not provide for strong inferences about implementation, as some have claimed that BOLD activity may arise from spurious correlations that do not reflect task performance (Laumann, et al., 2016) -. Shine (2021) proposed a neuroanatomically-inspired model for how cortical networks can shift their topologies (**Fig. 1**), whereby cortical shifts between integrated and segregated states result from influences of the basal ganglia (Bg) and cerebellum (Cb), respectively. The contributions of these two pathways is a result of the distinct thalamic subpopulations each project to, and the unique connection patterns these thalamic subpopulations have to the cortex, creating distinct functional pathways. According to this model, the basal ganglia flattens the energy landscape of cortical attractors, decreasing depth of basins and increasing the probable number of network states, thus exhibiting a more integrative topology. In contrast, the cerebellum deepens these basins, resulting in fewer cortical hub states and more modular network topologies (for a similar view, see Katsumi et al. (2021)). Although the Shine (2021) model is premised on the circuit logic of basal ganglia and cerebellar connections to thalamus and cortex, to our knowledge this subcortical control model has not been empirically tested at the macroscopic connectivity level using functional Magnetic Resonance Imaging (fMRI).

　　　Our goal in this work was to test the Shine (2021) model using a sample with a built-in replication test, where human participants completed modified versions of the Stroop (1935) and Multi-Source Inference (MSIT) tasks (Bush et al., 2003), while whole-brain hemodynamic responses were recorded with fMRI (full paradigm details available in Sheu et al., 2012). We relied on an edge time series approach that allows for estimating instantaneous functional connectivity matrices at each time sample (Faskowitz et al., 2020; Zamani Esfahlani et al., 2020). Based on the Shine (2021) model and prior work (Fair et al., 2007; Finc et al., 2017; Schultz & Cole, 2016), we hypothesized increased cognitive difficulty during the incongruent conditions of both tasks will cause cortical networks to shift to integrated states (i.e., low modularity). Conversely, the low cognitive effort during congruent conditions would result in segregated (modular) cortical networks. More importantly, the Shine (2021) model predicts increased basal ganglia connectivity to cortex will enable shifts of cortical networks to more integrated states, while cerebellar connections will influence the shift of cortical networks to more segregated topologies. The causal influence of these subcortical regions on cortical network topology is assessed using a temporal analysis of subcortical influence and cortical network modularity, to determine whether shifts in subcortical influence precede shifts in cortical topology.

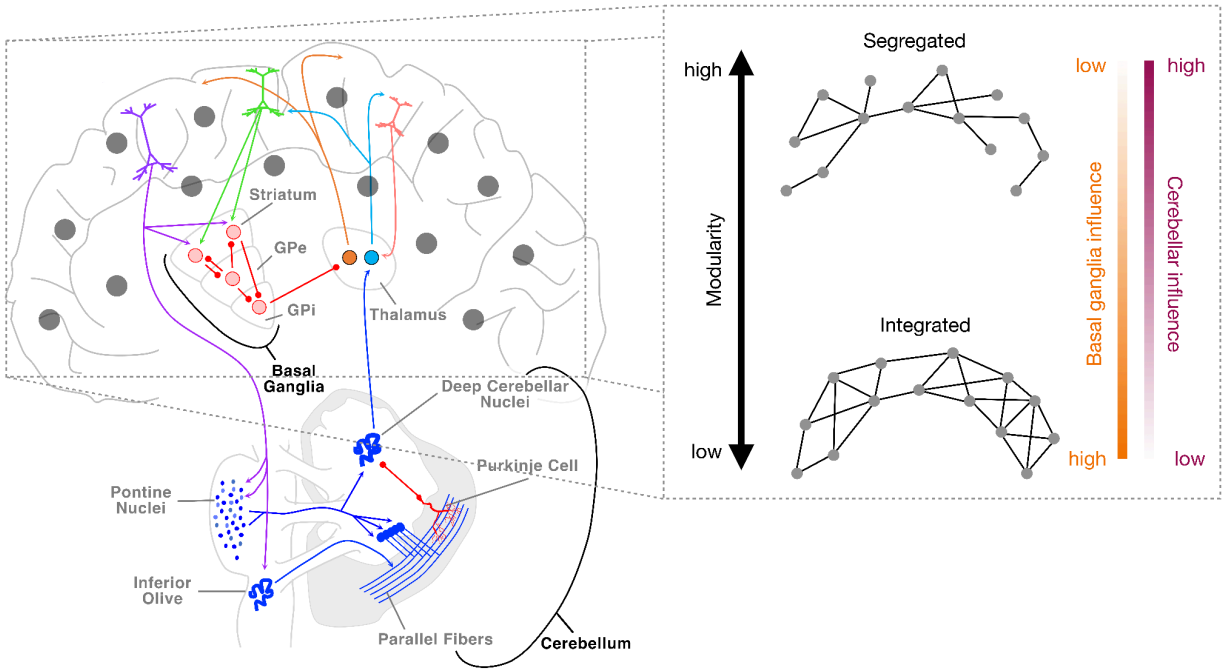

**Figure 1:** Illustration of anatomical connectivity of subcortical regions in the brain. The thalamus is used as a gating mechanism for signals relayed from globus pallidus interna (GPi) of the basal ganglia and the deep cerebellar nuclei of the cerebellum. Gray circles allude to hypothetical nodes (regions of interest) in a cortical network. Inset shows a high modularity index value indicates a more segregated network connectivity, while a low value an integrated network (black arrow). Orange line indicates eigenvector centrality and influence of the basal ganglia over the connectivity of the cortical network, while the purple line indicates influence of the cerebellum over the cortical network. **Fig. 1** adapted with permission from Shine (2021).

**METHODS**

<u>Participants</u>: A total of 242 midlife adults (119 identifying as female at birth, mean age = 40 ± 6 years, min age = 30, max age = 51 years) were included in this study, based on fMRI task and resting state data generated by the Pittsburgh Imaging Project (PIP). Selected scans for inclusion in the dataset - which entail three fMRI tasks - all have low average motion (mean framewise displacement (FD) was less than 0.35mm, as estimated in (Power et al., 2012)).

<u>Imaging Protocol</u>: MRI data was collected on a 3 Tesla Trio TIM whole-body scanner (Siemens, Erlangen, Germany), using a 12 channel head coil. For registration purposes a 1mm isotropic T1-weighted MPRAGE structural scan were acquired for each subject with the following specifications: repetition time (TR) = 2100ms, echo time (TE) = 3.29ms, field of view (FOV) = 256 x 208mm, inversion time = 1100ms, flip angle = $8°$. A 3mm isotropic functional blood oxygen level dependent (BOLD) scan was acquired with a T2*-weighted gradient echo-planar sequence using the following specifications:

repetition time (TR) = 2000ms, echo time (TE) = 28ms, field of view (FOV) = 205 x 205mm, flip angle = $90°$.

Task Acquisition: A variant of the Stroop (Stroop, 1935) and Multi-Source Inference tasks (MSIT; Bush et al., 2003) were used to assess neural response to congruent (no conflict) and incongruent (conflicting) information (Sheu, Jennings & Gianaros, 2012). Task blocks of congruence and incongruence were interleaved (**Fig. 2 B-C**) for a duration of 60s followed by a 10s delay where participants were instructed to fixate on a crosshair. In the congruent phase of the Stroop task participants were given a prompt of a color word visually shown with the associated color and the goal was to select the color the word specifies and was shown in. During the incongruent phase of this task participants were shown a prompt color word written in a different color from its spelling, and participants then had to select the color the word is presented in instead of the color the word specifies (Stroop, 1935). During the MSIT block participants were tasked with pressing one of three buttons to indicate an odd one out number. In the congruent phase the goal was to press the button corresponding to the location of the oddball number. While in the incongruent phase participants should suppress the urge to press the button associated with the location of the oddball number and instead press the button of the actual number value (Bush et al., 2003). As a control condition for difference in individual responses and to mediate an adaptive stress response on par with the performance of the subject, both the Stroop and MSIT incongruent task blocks included variable intertrial intervals. For the incongruent tasks blocks interval lengths were determined by the accuracy rating of the participant, where correct responses resulted in shortened rest periods to amplify the stressor conditions of the task and held accuracy at chance level. During the congruent blocks, trials were presented at the mean intertrial interval of the prior incongruent block, which resulted in approximately the same number of trials being presented for both congruent and incongruent blocks (Gianaros et al., 2017).

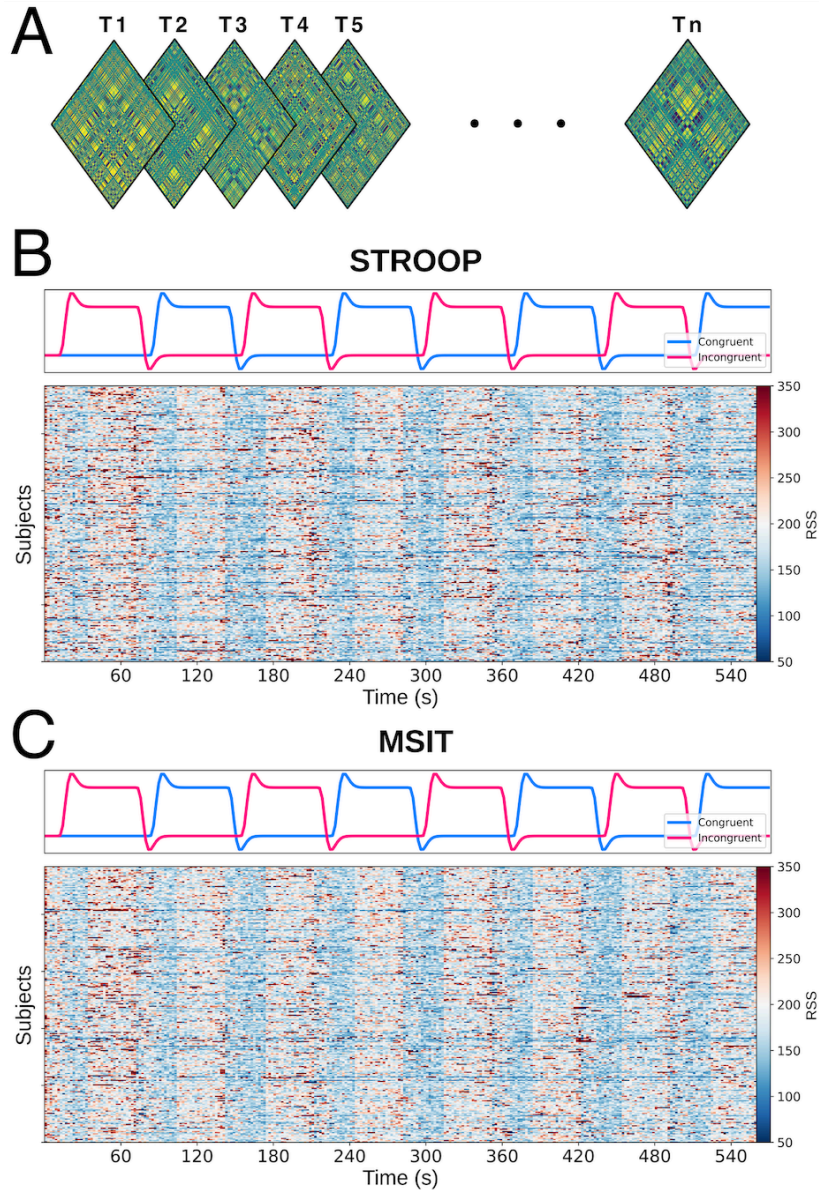

**Figure 2: (A)** Example matrices produced from edge time series, allowing for examination of network changes through the full temporal scale of an fMRI task. **(B-C)** Heatmap of all subjects (N=242) residual sum of squares (RSS) across the fMRI timescale on axis 2 (bottom), and is inherently the degree of coherent connectivity. Higher RSS on the colorbar indicates high integration of networks, while lower RSS values indicate lower integration of networks across the timescale. There is a clear separation between task blocks and fixation (lower RSS), as well as separation between incongruent blocks (higher RSS) from congruent blocks (lower RSS). Axis 1 (top) indicates the interleaved task block structure of incongruent and congruent Stroop trials. **Fig. 2 B-C** adapted from Rasero et al. (2023).

fMRI preprocessing: Data were preprocessed using the default fMRIprep pipeline (Esteban et al., 2018), an fMRI toolbox to aid scientists in consistency and reproducibly in preprocessing their data. This methodology performs anatomical registration, skull stripping and alignment and removal of physiological noise i.e. global signal regression (GSR) along with other steps, for the entirety of the fMRI timescale. These methods help decrease the possibility of false positives and negatives in the analysis pipeline, and produce a stronger signal through denoising confounds. Additionally, distortions of $B_0$ field inhomogeneity were corrected using the Fieldmap-less option in fMRIprep, which was implemented via nonlinear registration. Subsequently, to increase the signal-to-noise ratio a denoising step was performed. This involved regressing out 24 motion parameter (Friston et al., 1996) cosine terms accounting for oscillation effects greater than 187s, (average signal within white matter, cerebral spinal fluid (CSF), and whole brain tissue). Our denoising step also included the removal of the average activation signal for each task condition. Finally, the Shen et al. (2013) functional atlas, which includes the cortex, subcortex and cerebellum, was used to parcellate the brain into $k = 268$ functional regions, as alluded to in **Fig. 1** (gray dots).

Edge time series: Edge time series were calculated based on recent work from Faskowitz et al. (2020) and Zamani Esfahlani et al. (2021). Active networks were constructed along the timescale during congruent and incongruent phases, producing matrices at every time point (**Fig. 2A**). This was obtained by using, for example seeded regions $i$ and $j$, where initially the time series data is standardized to a *z*-score using equation 1. The pairwise dot product of all segmented brain regions is then calculated using equation 2 to derive the edge time series.

$$z_i = \frac{x_i - \mu_i}{\sigma_i} \tag{1}$$

$$Let\ c_{ij} = [z_i(1) \cdot z_j(1), \ldots, z_i(T) \cdot z_j(T)] \tag{2}$$

Modularity index: We used the Louvain algorithm (Brain Connectivity Toolbox for PYthon - bctpy) for maximizing the modularity function as it returns a scalar value, $Q$, signifying the modularity among cortical networks. More specifically, the modularity index was used to determine whether there is a high rate of segregation of cortical nodes during congruent time blocks with a converse effect observed during incongruent blocks. Prior to determining the modularity index, $Q$, the modularity matrix $B$ was determined; as noted in equation 3, where $A$ is the observed matrix of pairwise correlated moment-to-moment cortical activity and $P$ is the matrix of connections' expected correlation magnitudes, otherwise termed the null model which was used as standardization for the observed matrix $A$ (Zamani Esfahlani et al., 2021). Intuitively, communities detected using this null model correspond to groups of brain regions (or edges) whose connectivity during task blocks exceeds what is expected, under a degree preserving null model that is compatible with signed networks (Maslov & Sneppen, 2002; Rubinov & Sporns, 2011). In other words, the communities reflect groups of brain regions whose instantaneous co-fluctuation exceeds what would be expected under a signed degree-preserving network null model. At which point the calculation of the modularity matrix $B$ can be defined by equation 3, where the observed $A$ would be the functional task matrix and the expected $P$ would be a randomly permuted state matrix.

$$B = A - P \tag{3}$$

The modularity index $Q$ was then calculated using equation 4, where $\sigma \in \{1, \ldots, K\}$ and details which $K$ cortical community region $i$ or $j$ is segmented into $apriori$. $\delta(\sigma_i, \sigma_j)$ is the Kronecker function which returns 1 when $\sigma_i = \sigma_j$ and 0 otherwise, indicating whether two regions $i$ and $j$ are in the same community (Zamani Esfahlani et al., 2021). In the calculation of the signed version of the modularity index $Q$, $B_{ij}^+$ and $B_{ij}^-$ represent the separation of the signed aspects of the modularity matrix $B$, that is simply derived by thresholding the matrix to obtain two separate signed matrices (Rubinov & Sporns, 2011).

$$Q^{signed} = \sum_{ij} \left[ B_{ij}^+ - B_{ij}^- \right] \delta(\sigma_i, \sigma_j) \tag{4}$$

Eigenvector Centrality: We used eigenvector centrality (NetworkX) to investigate the influence of our subcortical target nodes, basal ganglia and cerebellum, to the rest of the cortical network. Subcortical nodes from the Shen et al. (2013) atlas were identified using a lookup table to associate each node with a specific brain region. This table was derived by manually examining different nodes of the atlas and identifying the various brain regions (credit is given to Andrew Gerlach for providing this table). Each subcortical region had multiple target nodes and thus eigenvector centrally values obtained from these nodes were averaged per region. Eigenvector centrality is a variant of traditional centrality, whereby it is a recursive summation of the centrality of the children of the target node. In this way the value takes into account not only how influential the target node children are but recursively how influential the subsequent children are to the cortical network. More formally, equation 5 highlights the summation for a target node $x_i$, where $A_{ij}$ is the associated adjacency matrix for child node $j$, $x_j$ is the eigenvector of the child node $j$ and $K_1^{-1}$ is the largest eigenvalue (Newman, 2010). In our case, $A_{ij}$ is the square matrix of edge time series from the Shen et al. (2013) atlas.

$$x_i = K_1^{-1} \sum_j A_{ij} x_j \tag{5}$$

Cortical configuration analysis: We did a within-subject examination of modularity values over the fMRI timescale using multiple linear regression with modularity as independent variable and the hemodynamic response function as dependent variables i.e. binarized values indicating task block. This examination provided $\beta$ coefficient values which indicated the direction of effect across the tasks. We also performed a test of mixed effects on the network state values to indicate whether there was a significant difference between the incongruent and congruent tasks, the blocks of the task or whether there was an interaction between the tasks and blocks of tasks. Subjects were used as random effects in the model. Modularity values were smoothed using gaussian filtering $(\sigma = 1)$ and *z*-scored prior to input into both regression models.

Influence of subcortical nodes: To investigate whether subcortical regions basal ganglia and cerebellum were controlling factors for shifts in cortical modularity we first collapsed each task condition into a single block by averaging their respective blocks across the fMRI timescale. This was done separately for each of our three variables (i.e. cortical modularity, basal ganglia and cerebellum eigenvector centrality), resulting in a vector of values including an incongruent block followed by a 10s fixation, and a congruent block with subsequent 10s fixation. We then conducted a cross correlation analysis using Pearson correlation (SciPy, Python), on each of our vector pairs for these three variables. We examined 30s forward (negative) and backward (positive) lags. The resulting patterns from the basal ganglia and cerebellum would determine whether the basal ganglia drives integration in the brain, while the cerebellum influences a segregated state (Shine, 2021).

Statistical analysis of data reliability To determine how efficacious our findings are replicated in the Stroop and MSIT datasets we conducted statistical tests for reliability. We sampled the correlation values at each lag and performed a one-sample t-test (SciPy, Python), with $popmean = 0$ to examine the strength of our obtained correlation values. P-values were then post hoc corrected using False Discovery Rate (FDR; SciPy, Python) and those under significance level lower than $0.05$ to be deemed significant.

Additionally, to test for between-dataset reliability, we conducted a bayes factor test, which quantifies the evidence for the alternative hypothesis relative to the null. Specifically, for each of our region of interest pairs (basal ganglia, cerebellum and cortex), we first selected the cross correlated values in the intersecting significant lags for Stroop and MSIT. Subsequently, the between-task Pearson correlation was determined for each of these overlapping significant lags and finally, based on this correlation coefficient, the bayes factor value was calculated (Pingouin, Python). In our case, the alternative hypothesis corresponded to correlation values greater than 0 (strong similarity), while the null corresponded to values equal to or lower than zero. In this way, bayes factors allowed us to determine to what extent MSIT was an appropriate validation dataset for our results obtained in the Stroop dataset. The bayes factor function, calculated using equation 25 of Ly, Verhagen & Wagenmakers (2016), is derived by integrating over the space of unknown parameters of the given datasets to determine the likelihood both datasets have similar parameters and can be used simultaneously.

**RESULTS**

**Task demand shifts cortical modularity.**

In our Stroop task, cortical modularity appears to be highly flexible across task conditions (**Fig. 3A,B**). Congruent blocks had overall higher modularity in comparison to incongruent blocks which had lower modularity. Since the modularity index metric is a proxy for indicating whether nodes are grouped in a modular fashion in the network, a higher score indicates more modularization (segregation), while a lower score indicates a more integrated state of the network. This observation was supported quantitatively as we obtained a mean regression coefficient ($\beta$) value of $0.072$, with $95\%$ confidence interval (CI) of $0.028, 0.116$ for incongruent task blocks. In congruent task blocks there was a $\beta$ coefficient of $0.128$ with $95\%$ CI of $0.081, 0.175$. The $\beta$ coefficients from the multiple linear regression model output allow us to determine direction of change in the modularity index values. In this instance there is a higher $\beta$ coefficient for modularity index values from congruent tasks blocks, indicating these blocks exhibit more modularization (segregation; **Fig. 3B Stroop**) in comparison to the incongruent task blocks which exhibit less modularization (integration; **Fig. 3B Stroop**). Using a linear mixed effects model, we found a significant difference in modularity between incongruent and congruent trial blocks for the Stroop task $(p = 0.014; std = 0.353; t = -2.464)$. There was no significant main effect of block $(p = 0.134; std = 0.089; t = -1.499)$ nor was the interaction between task and block significant $(p = 0.140; std = 0.125; t = 1.475)$. These results allow us to conclude there is lower modularity in the incongruent task blocks, which inversely indicates more integration of cortical networks during this task.

We next attempted to replicate our Stroop results using the MSIT dataset to examine validation effects of our findings and determine replicability of our observations. The multiple linear regression analysis for the incongruent condition resulted in a $\beta$ coefficient of $0.140$ with $95\%$ CI of $0.100, 0.113$ and in the congruent condition a $\beta$ coefficient of $0.155$ with $95\%$ CI of $0.113, 0.198$. In the mixed effects model we did not obtain a significant difference between incongruent and congruent conditions for MSIT $(p = 0.178; std = 0.353; t = -1.347)$. There was no significant main effect of block $(p = 0.266; std = 0.089; t = -1.114)$ nor was the interaction between task and block significant $(p = 0.172; std = 0.126; t = 1.365)$. Although we did not replicate our main effect of task condition on modularity in MSIT as in the Stroop task (Fig. 3D), this does not hinder investigation of our primary question of interest in terms of the subcortical control of cortical modularity as there is still substantial variation in modularity between fixation and task conditions in both Stroop and MSIT (Fig. 3A,C). This shift between task and fixation indicates there is a significant change in cortical network modularity, which occurs dependent on the performance of a task. The requirements of the MSIT did not produce similar network modularity results to Stroop task, even though they examine comparable conflict and no-conflict paradigms. These results support conclusions made from prior research (Rasero et al., 2023).

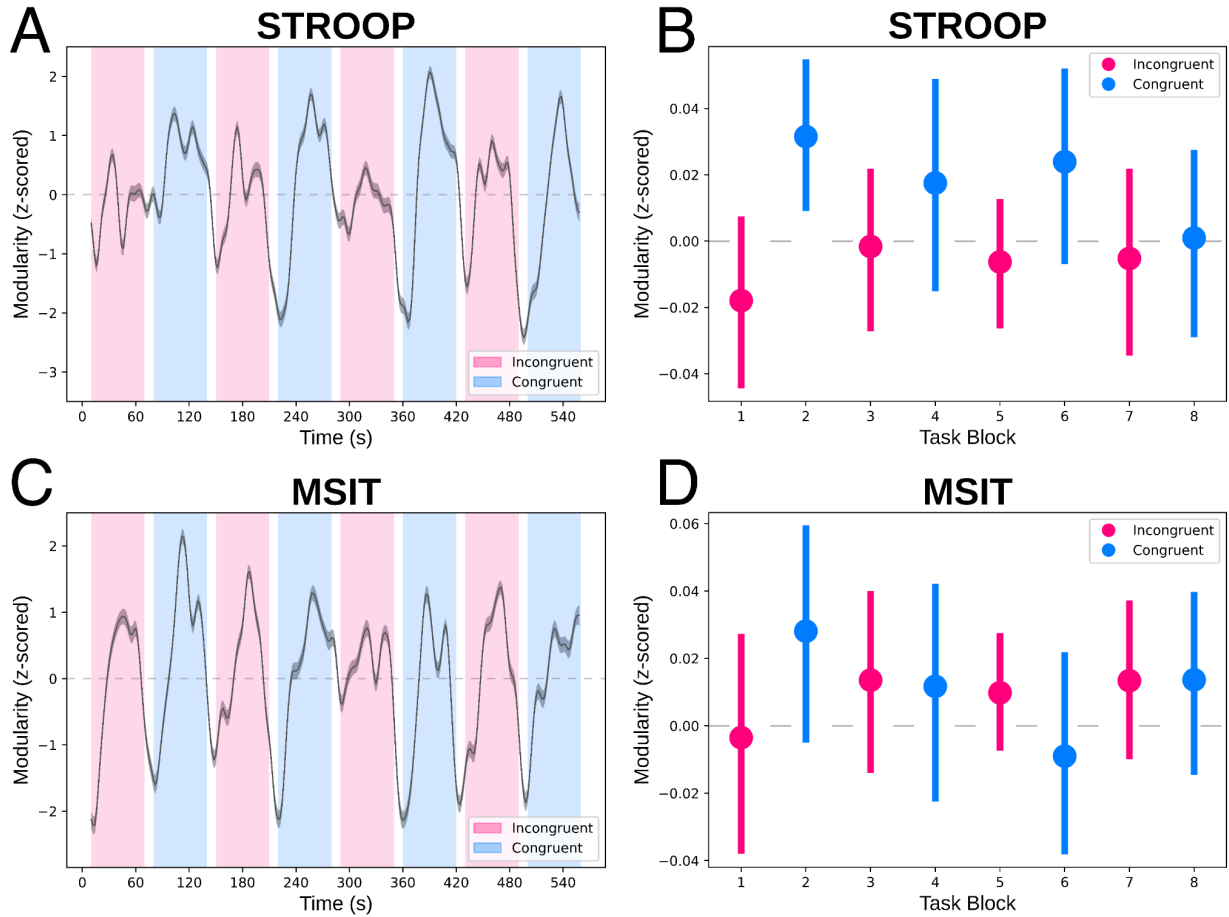

**Figure 3:** Z-scored cortical modularity index values were averaged across subjects and indicate shifts in network states across fMRI timescale. **(A, C)** Indicate cortical modularity values from the full fMRI timescale with error bars showing $95\%$ CI across subjects. **(A Stroop)** A significantly lower modularity (integration; $\beta = 0.072; 95\%$ CI = $0.028, 0.116$) is exhibited in incongruent task blocks with higher modularity (segregation; $\beta = 0.128; 95\%$ CI = $0.081, 0.175$) in congruent blocks. **(C MSIT)** We did not obtain a significant difference in cortical modularity values between task conditions for MSIT. **(B, D)** Indicate modularity values averaged by block with circles representing the mean and bars the $95\%$ confidence interval for modularity values across that block. **(B Stroop)** There is a distinct difference between cortical modularity during task conditions. **(D MSIT)** There is no observable difference between task conditions. Gaussian filtering $(\sigma = 1)$ was implemented to smooth temporal volatility.

### **Basal ganglia and cerebellum are differentially influential across tasks and rest.**

Eigenvector centrality is a metric to assess the influence a target node has on the rest of the network by assessing the recursive connectivity of the downstream children of that target node. For our hypothesis, this metric gives us an indication of how interconnected basal ganglia and cerebellum are to the cortical network over time. The score reflects the influence of subcortical circuits with cortex at each point in time. As shown in **Fig. 4**, we obtained a similar pattern of results for subcortical regions basal ganglia and cerebellum eigenvector centrality in both our Stroop and MSIT datasets, with a caveat of adjusted scale, i.e. the magnitude of the MSIT results are smaller than the Stroop.

There is an expectation based on the Shine (2021) model a more novel, cognitively demanding task will require greater integration of cortical network nodes. We hypothesized in the task blocks, subcortical node eigenvector centrality would reflect this difference in demand. Therefore in the more strenuous incongruent task basal ganglia nodes should exhibit more influence than they would in comparison to the congruent task and vice versa for the cerebellum i.e. more influence in the congruent task. Though the alternative was expected, we did not obtain a difference in the subcortical region eigenvector centrality across the task conditions. Instead we saw a consistent and general evoked response from task versus rest, represented here by the fixation blocks. In both tasks, the basal ganglia (**Fig. 4** orange line) exhibited the greatest influence in the beginning of the task, opposite to an initially low cortical modularity, followed by a subsequent dip in influence which coincided with a rise in cortical modularity. Low basal ganglia influence persisted until the 10s fixation between tasks, where it gradually increased to greatest influence for the beginning of the following tasks. This coordinated opposing activity of basal ganglia eigenvector centrality and cortical modularity could indicate the basal ganglia is working to drive the cortex into integration at optimal time points, as posited by the Shine (2021) model. Additionally, basal ganglia influence was greatest at the beginning of the task, for both incongruent and congruent conditions. This could indicate the basal ganglia is integral in the task regardless of difficulty.

Cerebellum (**Fig. 4** purple line) eigenvector centrality was generally anti-phasic compared to the basal ganglia influence, and followed instances of higher cortical modularity. For both Stroop and MSIT the cerebellar network had low influence in the beginning of the task, followed by a slight increase in influence mid task. At the end of the task block cerebellum influence was greatest and steeply decreased over the 10s fixation and rebounded to an ultimate low influence at the beginning of the next task block. The coordination of high cerebellar eigenvector centrality with high cortical modularity, particularly when they fluctuate together, could indicate the cerebellar networks are working to drive the cortex into a modular (segregated) network state, also aligned with the Shine (2021) model. The height of cerebellar activity at the end of the task for both incongruent and congruent task conditions could indicate the cerebellum is adept at task consolidation regardless of task difficulty.

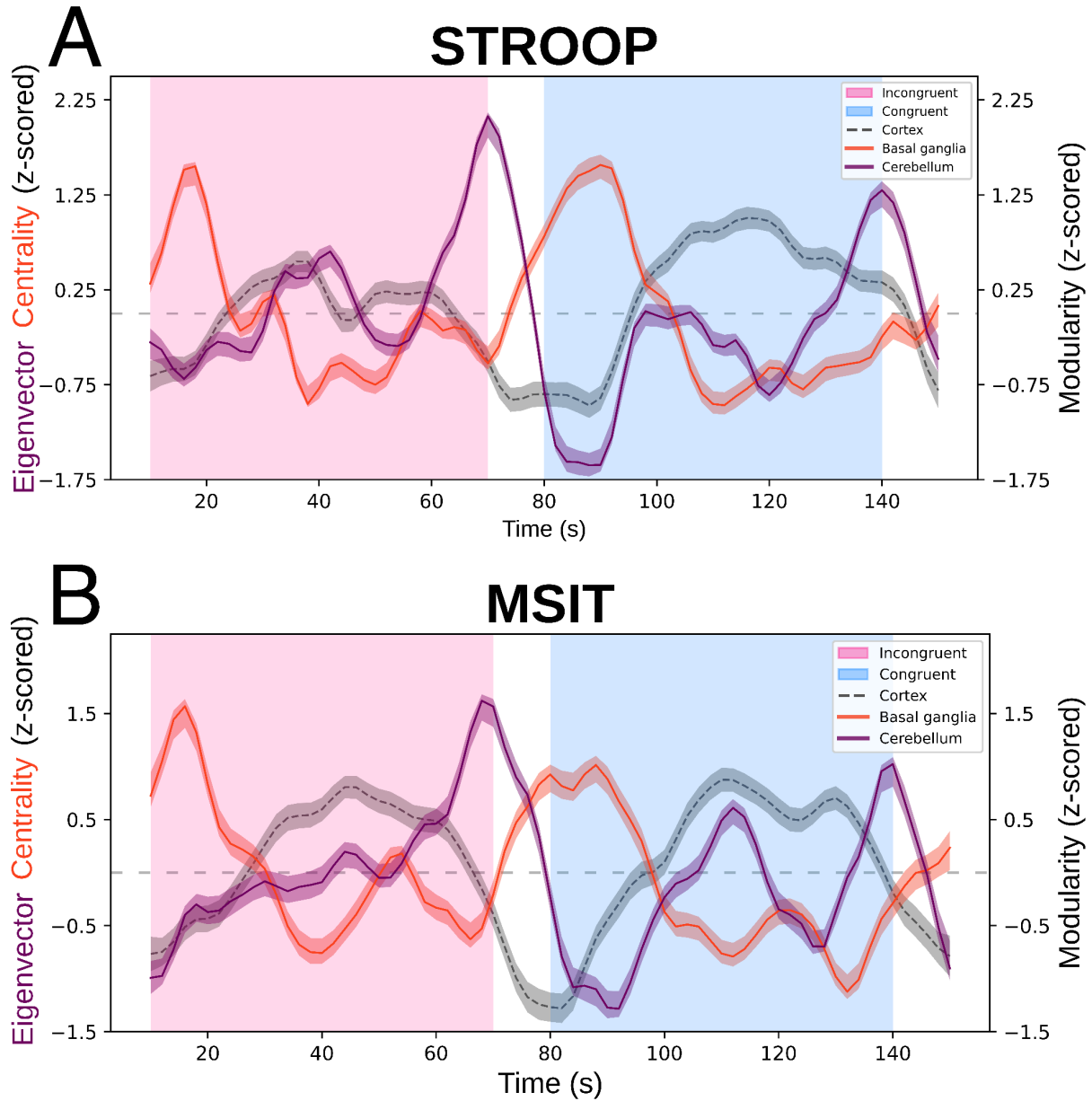

**Figure 4:** Z-scored cortical modularity index values (black), eigenvector centrality values for basal ganglia (orange) and cerebellum (purple) were averaged across subjects and indicate temporal shifts in network state across task conditions. Incongruent and congruent pairs of task conditions were averaged together to show a single representative sample of tasks across the fMRI timescale. Error bars for all regions of interest show $95\%$ CI across subjects. In both **(A)** Stroop and **(B)** MSIT we obtain similar results that are invariant of task condition. Basal ganglia (Bg) is more influential during the beginning of the task and decreases through the rest of the task. Bg influence then inclines again during the 10s fixation in preparation for the start of the following task block. Cerebellar (Cb) influence is inverse to that of basal ganglia during the task, there is initially low influence which is greatest towards the end of the task. Cb influence rapidly declines during the 10s fixation in preparation for observed low

influence at the start of the next task block. Gaussian filtering $(\sigma = 1)$ was implemented to smooth temporal volatility.

**Basal ganglia influence sets precedent for cortical network changes.**

We performed a cross correlation analysis, which lags two vectors forward (negative lag) and backwards (positive lag), to examine whether the subcortical influence precedes changes in cortical modularity or vice versa. Our results (**Fig. 5**) show the correlations of our three vectors (basal ganglia and cerebellar eigenvector centrality and cortical modularity) had temporal structure since our results are nonuniform. Our basal ganglia and cortex cross correlation analysis (**Fig. 5A**), showed the most significant correlations at negative lags -10, -8 and -6 seconds in Stroop (post hoc $p = 7.3951e - 04$) and -10, -8 lags in MSIT (post hoc $p = 0.0019$). This means changes in basal ganglia eigenvector centrality occur before changes in cortical modularity, consistent with a causal influence of the subcortical networks on cortical network topology. This effect was present in both the Stroop task, and with a lower magnitude, in the MSIT. This would be consistent with a phasic relationship between the subcortical and cortical networks.

In contrast, our main cerebellar and cortical cross correlation results (**Fig. 5B**) are at lag 0, due to this being the largest correlation value obtained for both Stroop (post hoc $p = 7.304e - 22$) and MSIT (post hoc $p = 6.04e - 29$). They share a positive correlation, indicating when cerebellar influence increases so does cortical modularity (**Fig. 4**). We did not observe a significant result in the negative lag direction (independent of the lag 0 correlation) that would indicate a possible causal role of cerebellar influence with cortical modularity. Instead, we see a prolonged, weak negative correlation in the positive lags appearing about 10-20 seconds (**Fig. 5B**), possibly reflecting feedback loops between the cerebellum and cortical network. Finally, we looked at the relationship between our subcortical networks themselves. At lag 0 for Stroop (post hoc $p = 6.857e - 09$) and MSIT (post hoc $p = 1.865e - 08$) the eigenvector centrality of basal ganglia and cerebellum (**Fig. 5C**) are negatively correlated with each other, as was qualitatively observed from their phasic dynamics in **Fig. 4**. Thus, as expected, the two subcortical networks work in opposite phases of each other.

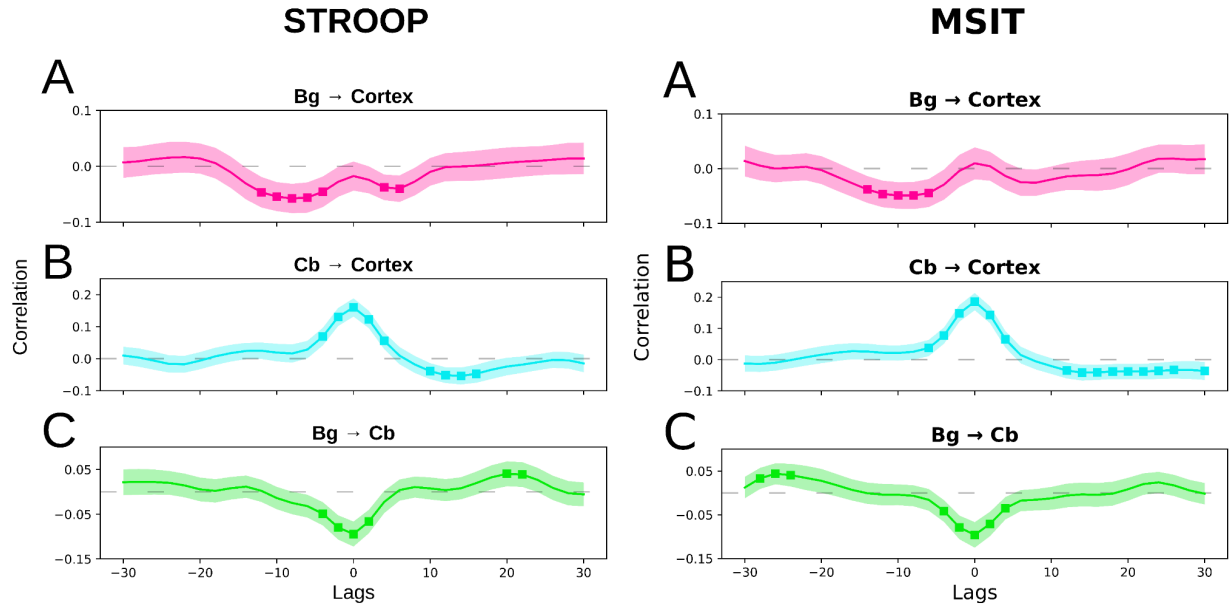

**Figure 5:** Pairwise cross correlation of forward (negative) and backward (positive) lags, using Pearson correlation of cortical modularity, basal ganglia and cerebellum eigenvector centrality. We obtained similar results for both the Stroop and MSIT datasets. **(A)** The forward lags for basal ganglia (Bg) and cortex highlighted the control Bg has on the resulting activity of the cortex, shown by the clear negative correlation. There was no significant correlation for the cortex exhibiting control on the Bg in MSIT, but there were a few significant lags that show this in Stroop. **(B)** Examination of the forward lags for cerebellum (Cb) cross correlated with the cortex in Stroop and MSIT did not show significant results. However, backward lags show the cortex influenced the resultant activity of the Cb. **(C Stroop)** There was no significant result for the forward lag, but the backward lag shows the Cb controls activity of the Bg. **(C MSIT)** There was no significant result for the backward lag, but the forward lag shows the Bg has control over activity of the Cb. Error bars for all regions of interest show $95\%$ CI across subjects. Squares indicate post hoc FDR corrected significant lag correlations.

### Inter-subject reliability between Stroop and MSIT datasets.

Our final test was a subject-wise examination of reliability between our Stroop and MSIT datasets, to determine whether they can be used interchangeably for validation purposes. We examined our Stroop cross correlation results (**Fig. 5 A-C**) and determined the most significant lags for basal ganglia and cortex (lags -10, -8 and -6), cerebellum and cortex (lag 0) and basal ganglia and cerebellum (lag 0) vector pairs. We then performed a Pearson correlation on the distribution of subject data between Stroop and MSIT. Basal ganglia and cortex had mostly weak and negative correlations between the two datasets at the significance lags ($r_{lag-10} = -0.0721, r_{lag-8} = -0.0857, r_{lag-6} = -0.0304$). In contrast, cerebellum and cortex ($r_{lag0} = 0.2157$) and basal ganglia and cerebellum ($r_{lag0} = 0.3087$) both had positive correlations. These results suggest there may be a difference in interchangeability of the datasets for validation purposes, dependent on the region of interest being examined.

We then used these correlation values to conduct a bayes factor analysis to quantify the evidence about the reliability of our datasets for use as equally representative models. In the event of dataset similarity meaningful positive correlation values should be obtained or alternatively they are different, in which case non-meaningful positive correlation values or negative correlation values should be obtained. Results for cerebellum and cortex ($bf_{lag-0} = 46.77$) and for basal ganglia and cerebellum ($bf_{lag-0} = 24193.322$) respectively indicate a strong and very strong evidence for an interchangeable purpose between both tasks. This does not occur for the basal ganglia and cortex, since their bayes factors provide negative evidence ($bf_{lag-10} = 0.0398, bf_{lag-8} = 0.036, bf_{lag-6} = 0.0575$).

**DISCUSSION**

The goal of this research was to examine cortical network modularity shifts in an fMRI task to evaluate the efficacy of the Shine (2021) model based on subcortical neuroanatomical pathways. The model hypothesizes basal ganglia and cerebellum are controlling mechanisms in the network transition from integration and segregation respectively, by relaying signals through the thalamus. Integrative states co-occur during cognitively strenuous tasks, while segregative states occur in simpler or habituated tasks. Using sample sets with built in replication, in the Stroop task we obtained a lower cortical modularity index (integration) during incongruent task blocks i.e. cognitively demanding, with comparable higher modularity (segregation) in congruent task blocks i.e. low cognitive demand. Our subcortical results indicated basal ganglia activity preceded cortical modularity, suggesting subcortical involvement in dynamical shifts of the cortex, and partial support for our hypothesis. We were not able to confirm the involvement of the cerebellum in initiating segregation of cortical networks. Peripheral results include correlative support for cortical undulation prior to influence of the cerebellum and prior cerebellar activity correlating with basal ganglia influence. Our results partially validate the Shine (2021) hypothesis as we did not show cerebellum has a preceding effect on cortical network segregation and only that basal ganglia has a preceding effect on integration. Ultimately, the theoretical approach of the Shine model may be more nuanced than initially proposed.

Findings of network shifts in cortical modularity from integration to segregation attributed to increased cognitive complexity, as validated in the Stroop task of this study, are commensurate with results obtained from previous studies. Lynn et al. (2021) examined and quantified detail balance (akin to segregation) and onset of entropy (integration) in the brain and found participants inherently maintain detail balance when they are not occupied with a task, but then shift to more entropic brain states as more cognitive load is required of them (Lynn et al., 2021). Similarly, Zhang & Saggar's (2022) research showed brain connectivity patterns might reflect multiple attractors (segregation) with overlapping connections and communication (termed bifurcations) rather than the operation of a single attractor in a dynamical landscape. This highlights the need for modular regions to communicate in a variety of conformations with other regions to effectively account for task completion, and most notably the need for integration during task difficulty in order to effectively complete task requirements. The need for integration of a brain system for effective learning is highlighted by Bassett et al. (2015) in their study where participants were tasked with learning a motor sequence tapping task of varying skill levels over a 6 week time course. At its core integration is necessary to solicit varying brain regions necessary to solve a task, and in this case the motor and visual networks exhibited heightened integration initially in skill learning that tampered to segregated networks over the time course. This is a great example of the need for integration as the task involves obvious motor movement, but visual regions are solicited as well to ensure effective precision in task completion. The study observed increased integration in other non-motor non-visual brain regions as well since this is a simplistic example of the regions necessary to solve the task e.g. hippocampal regions could be necessary for working memory and the cerebellum for maintaining task timing. As a task becomes more habitual there is less need to solicit these initially integrated brain regions for input as noted in the study.

     Though thalamic connections were not examined in this work they have been shown to be involved with change in cortical functional connectivity. Shine (2021) also hypothesized the thalamus would act as a control gating mechanism to funnel basal ganglia and cerebellar connections to the cortex. Müller et al. (2020) combined fMRI and calcium imaging to examine how core and matrix cells of the thalamus interact with the cortex to produce connectivity patterns, and found the matrix cells were responsible for integration and perpetuation of these signals long-term (Müller et al., 2020). Appropriate to our results of basal ganglia being a control state for cortical modularity, Ravizza & Ivry (2001) examined cerebellar and basal ganglia lesion patients and in a low motor task assessed their ability to shift between attentional states that constituted alternating between a distractor target task of varying difficulty. The authors found cerebellar lesion patients were able to effectively perform the attentional shift tasks while the basal ganglia lesion patients continued to have difficulty even though the motor aspects of the task were reduced and should have no effect on their performance. This study shows basal ganglia lesion participants in that study presumably were not able to effectively integrate cortical networks in order to meet task demands as the driver of this integrative state i.e. basal ganglia was disrupted in the participants' lesion (Ravizza & Ivry, 2001). This result is supported by another study, where participants who were more effective in controlling cortical oscillations, also showed more activity in the striatal region of their basal ganglia (Kasahara et al., 2022), highlighting its importance in facilitating integration.

     The limitations of this study entail the use of eigenvector centrality as a proxy for the control of subcortical nodes basal ganglia and cerebellum have on the cortical network. While this measure is a good assessment of how influential those nodes are to the rest of the connectivity of the cortical network, it is not a direct measure of control. We cannot be confident upon removal of the subcortical nodes the cortical network would fail to integrate and segregate as expected from our results. Another limitation is the use of cross correlation as this metric is an assessment of correlation of vector pairs of our regions of interest. Cross correlation does not imply causality or the activity of one region directly impacts another. An avenue of addressing these limitations would be

to take direct assessment of control by using network control theory measures mentioned in the following paragraph.

The first future direction of this work entails examining our hypotheses in *in silico* neural network models (similar to work by Ritz & Shenhav (2023)) to see whether the phenomenon of integration and segregation as a function of task difficulty is replicable. This would also give support to Shine (2021) hypothesis by determining whether brain networks settle into stable states, i.e. segregation, upon task habituation as a method of lowering the energy requirements of network conformational state changes. This aim would entail training a neural network to solve an incongruence task similar to Stroop and MSIT used in this study. We would then compute the modularity index scores to determine whether these *in silico* models have settled to integrative and segregative states when solving conflict and no-conflict tasks respectively similar to our experimental results in fMRI. This is similar to work by Ritz and Shenhav (2023) who had participants conduct a distractor and target task, whereby participants were trained to press specific buttons when specific colors crossed the screen, irrespective of dot motion direction. Cognitive control of the participants was assessed in this task as dots moving in the same direction as the selection arrow would constitute a congruence task, while those in the opposite direction an incongruence task. Additionally the experimenters varied dot coherence by different percentages to determine how participants reacted to distractors of different percentages. Lastly, experimenters varied the task by instructing participants to attend specifically to the dot color or the dot motion. Their results showed participants were primed for better performance given prior conflict and reward, indicating they became tuned to target recognition and distractor avoidance. Ritz and Shenhav then validated their experimental findings *in silico* by using an accumulator model to show that during attend-color trials the model had a faster reaction time with higher accuracy when color coherence increased, compared to when motion incongruence increased, where there was a longer reaction time and lower accuracy. This work is relevant to our intended future direction as it gives credence to the ability to validate experimental findings from conflict no-conflict tasks *in silico*.

The second future avenue of inquiry would be to determine whether the overall energy state of the network is greater in instances of integration in comparison to segregation, given that our neural network models exhibit integrative and segregative properties. The hypothesis behind this aim is that integrative states are helpful for promoting flexible task completion, however require more energy to maintain this state, which is why the goal of the network is to settle into a lower energy state of segregation. This aim would involve deriving a metric to effectively compute the energy of the network at a given time point and functional network connectivity state. The third future direction aims to determine whether, in a neural network with similar structural properties as the fMRI brain networks in our study - trained to solve a conflict no conflict task - the basal ganglia and cerebellar nodes would emerge as drivers nodes for state changes of the network. For this aim we would need to use a neural network configured with the connectomes of the brain, presumably a brain inspired graph neural network (GNN) would be appropriate for this task. One method to examine this aim would be to implement driver node detection or minimum necessary node algorithms to force change within a network (Nacher et al., 2019; Tang et al., 2012). Another method of analysis to determine controllability of the network is to examine how the subcortical regions basal ganglia and cerebellum network co-fluctuations resemble those of cortical networks over time and determine instances and subnetworks where they are similar (Betzel et al., 2022).

In conclusion, this work sought to examine the relationship between network state transition in the brain dependent on a simultaneously performed task. We found during the incongruent phases of the Stroop task there is significant integration of nodes across the network. Whereas, during the congruent phases of the task there is more segregation of networks. These results imply in order to perform a more cognitively strenuous task the brain seeks to employ across network communication to efficiently compute the requirements of the task. However, when doing a simpler task as in the congruent phase of the experiment, the brain does not require this heightened cross network communicative state and can instead solve the requirements of the task by communicating within network community hubs. The cortical results of this study build on prior work that have exhibited this integrative and segregative phenomenon in the brain (Wang et al., 2021; Cole et al., 2013) and are a confirmation of the first portion of our hypothesis.

Our second aim for this work was to examine whether the subcortical regions of the basal ganglia and the cerebellum are influential in the control of these cortical network reconfigurations. We qualitatively assessed these subcortical regions by examining the eigenvector centrality for basal ganglia and cerebellum with cortical modularity (**Fig. 4**). We determined the subcortical regions basal ganglia and cerebellum were negatively correlated throughout the timescale, indicating interchanging activity (**Fig. 5**). The cerebellum was positively correlated with cortical modularity, presumably indicating a rise in activity with a coordinated increase in cortical modularity. Using cross correlation we determined the basal ganglia activity has a control effect on cortical modularity. We were not able to determine this for the cerebellum. Additionally we obtained results indicating the cerebellum has a control effect on basal ganglia. Overall, the basal ganglia exhibited more influence in the beginning of the task (possibly as an initiator) while the cerebellum at the end (potentially for consolidation). It is our belief, we obtained partial support for our hypothesis that subcortical regions act as control states for cortical network reconfiguration. Additionally, while the Shine (2021) theory has merit in this respect, we believe it may ultimately be more nuanced in its interpretation.

## RESOURCES

The [pre-registration](#) for this project can be found on OSF Registries with the following title: Reconfiguration of cortical network connectivity with increasing cognitive complexity. [Code](#) used in this paper is available on GitHub and data is uploaded on [KiltHub](#).

## ACKNOWLEDGEMENTS

Thanks are extended to James Mac Shine for providing the original image used in **Fig. 1** and permission to make edits for this paper and Andrew Gerlach for providing a list of brain regions associated with each node of the Shen et al. (2013) atlas.

**FUNDING**

This work was supported by the National Institutes of Health P01HL040962 and R01089850.